# Applicability of molecular statics simulation to partial dislocations in GaAs

Thomas Riedl[*,a,b] and Jörg K.N. Lindner[a,b]

[a] Department of Physics, University of Paderborn, Warburger Straße 100, 33098 Paderborn, Germany
[b] Center for Optoelectronics and Photonics Paderborn (CeOPP), Warburger Straße 100, 33098 Paderborn, Germany

**Abstract**

The suitability of molecular statics (MS) simulations to model the structure of 90° glide set partial dislocation cores in GaAs is analyzed. In the MS simulations the atomic positions are iteratively relaxed by energy minimization, for which a Tersoff potential parametrization appropriate for nanostructures has been used. We show that for the Ga terminated partial the resulting bond lengths of the atoms in the dislocation core agree within 5-10% with those of previous density functional theory studies, whereas a significant discrepancy appears in the case of the As terminated partial.

## 1 Introduction

Dislocations are frequently observed one-dimensional crystal defects, which arise in the course of crystal growth, mechanical deformation and ion implantation. On the one hand, dislocations are accompanied with deep electronic states leading to non-radiative carrier recombination, which limits the performance of semiconductor devices [1]. On the other hand, dislocations can be utilized to create Si-based light emitting devices [2], and to getter impurity atoms [3].

The structure of the dislocation core determines the dislocation properties such as line energy, mobility and electronic behavior. Atomistic modeling of the core structure has been realized using (i) empirical potentials such as the valence force field method [4,5] and (ii) ab-initio density functional theory (DFT) [6-8]. The latter enables high accuracy, but is restricted to relatively small systems (typically few hundreds of atoms). In contrast, molecular statics / molecular dynamics (MS/MD) simulations based on empirical potentials offer larger computational speeds, which permit to model both the dislocation core and the far-reaching strain field. This is of particular interest in the case of nanostructures, where continuum elasticity fails to describe the core structure and DFT is computationally not feasible for the entire nanostructure. In semiconductor nanowire heterostructures, which are highly attractive for optoelectronic and nanoelectronic applications, dislocations can occur if the critical dimensions for misfit defect formation are exceeded [9]. Therefore, the application of MS/MD to such nanostructures is desirable in order to analyze the dislocation stability and dynamics as well as their strain fields. However, for this task suitable potential parametrizations are needed, which enable reliable modelling of surfaces and defects. For GaAs and InAs a number of parametrizations of the Tersoff bond order potential [10] have been proposed [11-18]. While most of them are fitted only to bulk and in some cases to dimer properties, the parametrization by Hammerschmidt et al. predicts bulk and surface properties of GaAs and InAs reasonably well [18]. This parametrization also reproduces the formation energies of vacancies in GaAs within 10%. However, the formation energies of antisite defects, especially the $As_{Ga}$ antisite, are not correctly predicted [18]. The fourfold As coordinated As atom in this defect conflicts with the choice of the potential parameters, which are optimized for As dimer reconstructed surfaces [18]. As reconstructed dislocation cores involve bonds between atoms of the same chemical element, the question arises with which accuracy the structure and energy of dislocation cores can be modelled using this Tersoff parametrization.

---

[*] Corresponding author: e-mail thomas.riedl@uni-paderborn.de, Phone: +49 5251 60 2746, Fax: +49 5251 60 3247

In the present study we analyze the applicability of Tersoff potential based MS simulations to partial dislocation cores in GaAs by comparing with the results of published DFT studies. In the case of a reliable description, MS/MD using empirical potentials could be applied to dislocations in nanostructures for computing their energy as well as the energy barriers during dislocation motion.

## 2 Calculation

The considered dislocation configurations comprise alpha and beta 90° partials of the glide set in zinc-blende GaAs with Burgers vector $a/6[\bar{1}1\bar{2}]$ and line direction $[110]$, the same as in previous DFT studies [19-21]. In the case of glide set dislocations with their cores situated between the pairs of narrowly spaced $\{111\}$ planes, alpha or beta means that the dislocation line is terminated by the negatively (positively) charged element, here As or Ga. In order to approximate a bulk environment, the dislocation was placed in the centre of a disc-shaped simulation cell with height $\sqrt{2}a$, radius 140 $a/\sqrt{2}$ ($a$: lattice parameter) and periodic boundaries at the bottom $(\bar{1}\bar{1}0)$ and top $(110)$ disc faces (Fig. 1). Due to the large disc radius the interaction of the dislocation strain field with the non-periodic radial surfaces can be neglected. In the initial state the atoms close to the dislocation core were arranged according to the unreconstructed state and displaced according to the dislocation strain field given by continuum elasticity theory [22].

In the MS simulation the lattice was iteratively relaxed by energy minimization based on the conjugate gradient method using the LAMMPS software [23]. For the atomic interactions the Tersoff potential was employed [10], which has been proven suitable for III-V semiconductors [12-16]. In this approach the total energy of the system is composed of a repulsive and an attractive component, where the latter is modified by a bond order term taking the dependence of bond strength on the coordination number into account. As parametrization we adopted that of Hammerschmidt et al., since it was shown to yield the most accurate description of GaAs and InAs bulk and surface properties [18]. As stopping criterion for the energy minimization an energy tolerance, i.e. a relative energy difference between successive iterations of $10^{-9}$ was chosen.

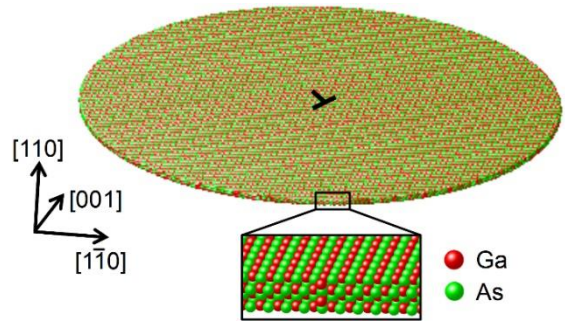

**Figure 1** Oblique view of the simulation cell. The position of the dislocation is indicated by the black symbol. The enlargement illustrates the crystalline arrangement of the Ga and As atoms.

Atom coordinate lists required as input for the LAMMPS software were calculated by using self-written scripts in the DigitalMicrograph software [24]. Evaluation of atom positions after relaxation was also performed with DigitalMicrograph scripts.

## 3 Results and Discussion

Fig. 2 depicts the relaxed structure of the 90° As terminated (alpha) partial dislocation core. As the dangling bonds in the core of this glide set dislocation are nearly parallel to the slip plane, a bond reconstruction occurs, leading to the formation of bonds between chemically identical core atoms. The alpha partial displays a strong double period (DP) reconstruction (Fig. 2b), i.e. alternating shorter and longer bonds appear between the As core atoms.

Each As core atom is connected with four nearest-neighbour Ga atoms. Three of these bonds (1-3 in Fig. 2b) form shallow angles with the slip plane, the remaining one (4 in Fig. 2b) is directed clearly out of the slip plane. Table 1 compares the resulting lengths of the bonds involving core atoms with those reported in the literature. In our case the As-As bonds of the DP reconstructed alpha core (marked by * and ** in Fig. 2b) have lengths of 1.93 Å and 2.64 Å. In contrast, previous DFT studies report As-As bond lengths of 2.60 Å [19,20] and 2.69 Å [21] for the single period (SP) reconstruction, and 2.60 Å, 2.64 Å for the DP reconstruction [21]. Based on the sum of atomic radii [25] an As-As bond length of 2.30 Å is estimated. For better comparison with our results all literature bond lengths from DFT have been scaled in order to correct for the difference in the equilibrium lattice parameter between MS and DFT calculations. Clearly, the MS simulation predicts a larger difference between the DP reconstructed bond lengths and a shorter average length of these bonds. The discrepancy between the DFT and our results can be ascribed to the inability of the Hammerschmidt parametrization used in the MS simulation to describe the As-As bonds correctly.

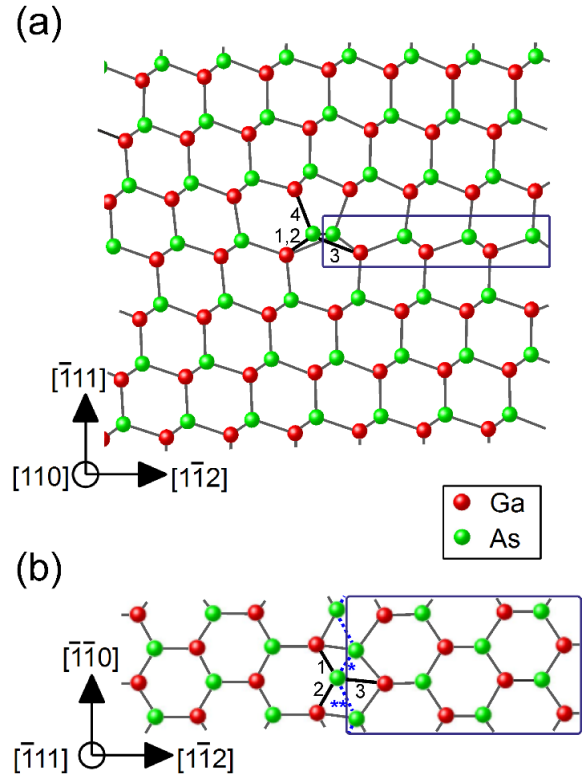

**Figure 2** Atomistic model of a 90° alpha partial dislocation in GaAs, obtained by molecular statics relaxation. (a) View along the $[110]$ dislocation line, (b) view onto the $(\bar{1}11)$ glide plane. In (b) bonds between As core atoms are drawn as blue dotted lines and marked by * and **. In (a) and (b) bonds between one selected As core atom and nearest-neighbour Ga atoms are numbered with 1, 2, 3, 4, respectively. The rectangle marks the region of the stacking fault.

**Table 1** Bond lengths in Å of the atoms in the core of the 90° alpha partial in GaAs.

|  | Present study (MS) | Local DFT cluster method [19,20][1] | DFT local density approximation [21][2] |
|---|---|---|---|
| As-As core bonds | 1.93, 2.64 | 2.60 | SP: 2.69; DP: 2.60, 2.64 |
| As-Ga backbonds | Parallel to slip plane: 2.69, 2.74, 2.82<br>Normal to slip plane: 2.62<br>Average: 2.72 | Parallel to slip plane: 2.56, 2.44, 2.65, 2.45<br>Normal to slip plane: 2.41, 2.42<br>Average: 2.49 | SP: Range: 2.40-2.69[3]<br>Average: 2.45[3]<br>DP: Range: 2.41-2.63[3]<br>Average: 2.45[3] |

[1] Original values scaled with factor 5.653:5.61
[2] Original values scaled with factor 5.653:5.591
[3] Bond lengths within 7 Å of the dislocation core

The average length of the backbonds between As core atoms and adjacent Ga atoms is 2.72 Å, which exceeds the unstrained Ga-As bond length in GaAs of 2.45 Å by 11%, due to the higher atomic coordination and strain in the dislocation core. Sitch et al. [19] and Öberg et al. [20] found an average As-Ga backbond length of 2.49 Å (8% smaller than MS, but still larger than in unstrained GaAs), whereas for a region within 7 Å of the core Beckman and Chrzan report an average bond length of 2.45 Å [21], only marginally larger than in unstrained GaAs.

The relaxed core structure of the 90° Ga terminated (beta) partial is illustrated in Fig. 3. Table 2 compares the bond lengths resulting from MS with literature values. The MS simulation yields an SP reconstructed beta core with a Ga-Ga bond length of 2.31 Å. Previous DFT studies report 2.43 Å [19,20] (5% larger) and 2.53 Å (10% larger) for the SP reconstruction, and 2.48 Å, 2.49 Å for the DP reconstruction [21] (7-8% larger). The sum of atomic radii [25] yields a Ga-Ga bond length estimate of 2.60 Å.

Similarly to the alpha core, the average length of the backbonds between Ga core atoms and adjacent As atoms of 2.61 Å in the beta core is 7% larger than the unstrained Ga-As bond length in GaAs. Sitch et al. [19] and Öberg et al. [20] report an average Ga-As backbond length of 2.46 Å, i.e. 6% smaller than MS. Within 7 Å of the core, according to Beckman and Chrzan, the average bond length amounts to 2.45 Å, very close to the Ga-As bond length in unstrained GaAs [21].

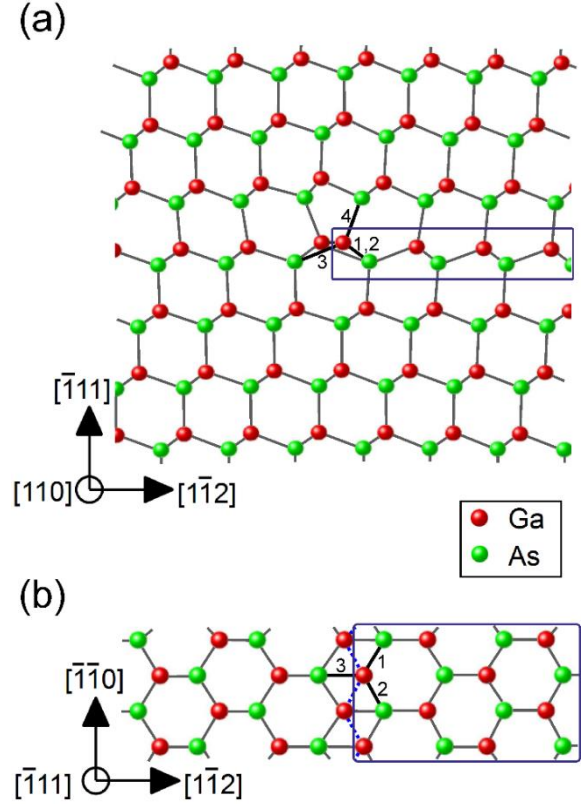

**Figure 3** Atomistic model of a 90° beta partial dislocation in GaAs, obtained by molecular statics relaxation. (a) View along the $[110]$ dislocation line and (b) view onto the $(\bar{1}11)$ glide plane. In (b) bonds between Ga core atoms are drawn as blue dotted lines. In (a) and (b) bonds between one selected Ga core atom and nearest-neighbour As atoms are numbered with 1, 2, 3, 4, respectively. The rectangle marks the region of the stacking fault.

**Table 2** Bond lengths in Å of the atoms in the core of the 90° beta partial in GaAs.

|  | Present study | Local DFT cluster method [19,20][1] | DFT local density approximation [21][2] |
|---|---|---|---|
| Ga-Ga core bonds | 2.31 | 2.43 | SP: 2.53; DP: 2.48, 2.49 |
| Ga-As backbonds | Parallel to slip plane: 2.61, 2.61, 2.69<br>Normal to slip plane: 2.54<br>Average: 2.61 | Parallel to slip plane: 2.50, 2.45, 2.56, 2.49<br>Normal to slip plane: 2.40, 2.39<br>Average: 2.46 | SP: Range: 2.41-2.56[3]<br>Average: 2.45[3]<br>DP: Range: 2.40-2.59[3]<br>Average: 2.45[3] |

[1]Original values scaled with factor 5.653:5.61
[2]Original values scaled with factor 5.653:5.591
[3]Bond lengths within 7 Å of the dislocation core

## 4 Conclusions

In summary, we have evaluated the applicability of atomistic MS simulations to glide set 90° partial dislocations in GaAs by comparing with the results of previous DFT studies. We find that the Hammerschmidt parametrization of the Tersoff potential enables a reasonable MS description of the group III (Ga) terminated 90° partial. The bond lengths of the atoms in the dislocation core obtained by MS agree with those of the DFT studies within 5-10%. In case of the group V (As) terminated partial the agreement with DFT is worse, owing to the imprecision in the description of the As-As interaction different from that in surface dimers.

**Declarations of interest:** none

**Acknowledgements**

This work was supported by the Deutsche Forschungsgemeinschaft [grant numbers RI 2655/1-1, LI 449/16-1].

**References**

[1] S. Mahajan, Acta Mater. 48 (2000) 137-149. https://doi.org/10.1016/S1359-6454(99)00292-X

[2] M. Kittler, M. Reiche, T. Arguirov, W. Seifert, X. Yu, Phys. Status Solidi 203 (2006) 802. https://doi.org/10.1002/pssa.200564518

[3] J. Lu, X. Yu, Y. Park, G. Rozgonyi, J. Appl. Phys. 105 (2009) 073712. https://doi.org/10.1063/1.3093912

[4] M. S. Duesbery, B. Joos, D. J. Michel, Phys. Rev. B 43 (1991) 5143-5146. https://doi.org/10.1103/PhysRevB.43.5143

[5] S. Marklund, Wang Yong-Liang, Solid State Comm. 91 (1994) 301-305. https://doi.org/10.1016/0038-1098(94)90306-9

[6] J.-S. Park, J. Kang, J.-H. Yang, W. E. McMahon, S.-H. Wei, J. Appl. Phys. 119 (2016) 045706. https://doi.org/10.1063/1.4940743

[7] K. E. Kweon, D. Åberg, V. Lordi, Phys. Rev. B 93 (2016) 174109. https://doi.org/10.1103/PhysRevB.93.174109

[8] I. Belabbas, J. Chen, M. I. Heggie, C. D. Latham, M. J. Rayson, P. R. Briddon, G. Nouet, Modelling Simul. Mater. Sci. Eng. 24 (2016) 075001. https://doi.org/10.1088/0965-0393/24/7/075001

[9] H. Ye, P. Lu, Z. Yu, Y. Song, D. Wang, S. Wang, Nano Lett. 9 (2009) 1921-1925. https://doi.org/10.1021/nl900055x

[10] J. Tersoff, Phys. Rev. B 37 (1988) 6991-7000. https://doi.org/10.1103/PhysRevB.37.6991

[11] R. Smith, Nucl. Instr. Methods Phys. Res. B 67 (1992) 335-339. https://doi.org/10.1016/0168-583X(92)95829-G

[12] P. A. Ashu, J. H. Jefferson, A. G. Cullis, W. E. Hagston, C. R. Whitehouse, J. Crystal Growth 150 (1995) 176-179. https://doi.org/10.1016/0022-0248(95)80202-N

[13] K. Albe, K. Nordlund, J. Nord, A. Kuronen, Phys. Rev. B 66 (2002) 035205. https://doi.org/10.1103/PhysRevB.66.035205

[14] M. A. Migliorato, A. G. Cullis, M. Fearn, J. H. Jefferson, Phys. Rev. B 65 (2002) 115316. https://doi.org/10.1103/PhysRevB.65.115316

[15] J. Nord, K. Albe, P. Erhart, K. Nordlund, J. Phys.: Condens. Matter 15 (2003) 5649-5662. https://doi.org/10.1088/0953-8984/15/32/324

[16] D. Powell, M. A. Migliorato, A. G. Cullis, Phys. Rev. B 75 (2007) 115202. https://doi.org/10.1103/PhysRevB.75.115202

[17] J. T. Titantah, D. Lamoen, M. Schowalter, A. Rosenauer, J. Appl. Phys. 101 (2007) 123508. https://doi.org/10.1063/1.2748338

[18] T. Hammerschmidt, P. Kratzer, M. Scheffler, Phys. Rev. B 77 (2008) 235303. https://doi.org/10.1103/PhysRevB.77.235303

[19] P. Sitch, R. Jones, S. Öberg, M. I. Heggie, Phys. Rev. B 50 (1994) 17717-17720. https://doi.org/10.1103/PhysRevB.50.17717

[20] S. Öberg, P. K. Sitch, R. Jones, M. I. Heggie, Phys. Rev. B 51 (1995) 13138-13145. https://doi.org/10.1103/PhysRevB.51.13138

[21] S. P. Beckman, D. C. Chrzan, Phys. Status Solidi B 243 (2006) 2122-2132. https://doi.org/10.1002/pssb.200666808

[22] J. P. Hirth, J. Lothe, Theory of Dislocations, second ed., Wiley, New York, 1982.

[23] S. Plimpton, J. Comp. Phys. 117 (1995) 1-19. https://doi.org/10.1006/jcph.1995.1039

[24] Gatan Inc., Gatan Microscopy Suite Software. http://www.gatan.com/products/tem-analysis/gatan-microscopy-suite-software (accessed 16 May 2019).

[25] J. C. Slater, Quantum Theory of Molecules and Solids, Vol. 2, first ed., McGraw Hill, New York, 1965.